\documentclass[floats,twocolumn,pra]{revtex4}
\usepackage{amssymb}
\usepackage{graphicx}
\usepackage{amsmath}
\usepackage{amsfonts}
\usepackage{accents}
\usepackage{color}
\usepackage[normalem]{ulem}
\newtheorem{theorem}{Theorem}

\newtheorem{corollary}[theorem]{Corollary}

\begin{document}
\title{Bounds for multi-end communication over quantum networks}
\author{Stefano Pirandola}
\affiliation{Department of Computer Science, University of York, York YO10 5GH, United Kingdom}
\affiliation{Research Laboratory of Electronics, Massachusetts Institute of Technology,
Cambridge, Massachusetts 02139, USA}

\begin{abstract}
Quantum and private communications are affected by a fundamental limitation
which severely restricts the optimal rates that are achievable by two distant
parties. To overcome this problem, one needs to introduce quantum repeaters
and, more generally, quantum communication networks. Within a quantum network,
other problems and features may appear when we move from the basic unicast
setting of single-sender/single-receiver to more complex multiend scenarios,
where multiple senders and multiple receivers simultaneously use the network
to communicate. Assuming various configurations, including multiple-unicast,
multicast, and multiple-multicast communication, we bound the optimal rates
for transmitting quantum information, distributing entanglement, or generating
secret keys in quantum networks connected by arbitrary quantum channels. These
bounds cannot be surpassed by the most general adaptive protocols of quantum
network communication.

\end{abstract}
\maketitle

\section{Introduction}

Quantum and private communications represent some of the most advanced areas
of quantum information~\cite{NiCh,first,HolevoBOOK,review,BraRMP}. In
particular, quantum key distribution (QKD)~\cite{BB84,Ekert} has been already
developed into several commercial prototypes, besides the fact that
quantum-secured networks and satellite quantum communications are being
developed by various countries~\cite{QKDadvance}. A more ambitious and
long-term goal is that of the quantum
internet~\cite{Kimble,HybridINTERNET,Whener} where remote quantum computers
are suitable connected by optical links so as to ultimately create a worldwide
architecture for distributed quantum computing.

In terms of quantum communications, one of the basic reasons to build quantum
networks~\cite{Rod} is to overcome the rate limitations of point-to-point
protocols. As shown in Ref.~\cite{QKDpaper}, the maximum rates at which two
remote parties can transmit quantum information, distribute entanglement or
secret correlations over a lossy channel of transmissivity $\eta$ are all
equal to $-\log_{2}(1-\eta)$, also known as the
Pirandola-Laurenza-Ottaviani-Banchi (PLOB) bound. For the specific case of
QKD, this ultimate point-to-point rate can be achieved by employing a quantum
memory at the sender side, as shown back in 2009 when the notion of reverse
coherent information was introduced for bosonic
channels~\cite{ReverseCAP,RevCohINFO}. On the other hand, if a middle node is
inserted between the remote parties, the PLOB bound can be practically beaten,
as shown by the recent twin-field QKD\ protocol~\cite{Marco}.

Once understood that the use of relays or
repeaters~\cite{Briegel,Rep2,Rep3} can overcome the PLOB bound, it
is also important to understand the ultimate limits achievable by
repeater-assisted quantum
communications~\cite{netpaper,longVersion}. Using techniques from
network information
theory~\cite{Slepian,Schrijver,Gamal,Cover,netflow} and very
recent channel simulation tools developed in quantum
information~\cite{QKDpaper}\ (see also
Refs.~\cite{Metro,nonPauli,TQC,BK2,Qmetro,revSENS}), one can bound
or exactly derive the capacities for quantum and private
communication between two end-points of a repeater chain or a
quantum network. This was shown in Ref.~\cite{netpaper} which
reports the end-to-end (unicast) results originally established in
the 2016 unpublished work~\cite{longVersion}.


The present paper reports and refines the other (multi-end)
results of unpublished~\cite{longVersion}, thus providing a
generalization of Ref.~\cite{netpaper} from the unicast setting of
single-sender/single-receiver to more complex scenarios where
multiple senders and receivers are involved. In these scenarios,
the remote parties compete with the others in order to make an
optimal use of the quantum network.
We assume different configurations, including multiple-unicast
(where there are many single-sender/single-receiver pairs trying
to communicate in a simultaneous fashion), multicast (where a
single-sender tries to communicate with multiple receivers), and
multiple-multicast (where different senders try to communicate
with the same set of multiple receivers). In all these
communication configurations, we derive single-letter upper bounds
for the maximum rates at which the parties can transmit quantum
information, distribute entanglement or secret keys. These bounds
are valid for networks connected by arbitrary quantum channels and
are expressed in terms of the relative entropy of entanglement
(REE)~\cite{RMPrelent,VedFORMm,Pleniom}.

It is important to remark that the results apply to both discrete-
and continuous-variable systems, i.e., quantum networks connected
by quantum channels acting over finite- or infinite-dimensional
spaces. In fact, as discussed afterwards, the theory presented for
finite dimension $d$ can be extended to $d = +\infty$ by
generalizing the notion of channel simulation to an asymptotic
formulation which is based on a sequence of finite-energy resource
states. Thanks to this tool, we can compute the relevant
functionals on the sequence and then take the infinite-energy
limit of their values over the sequence. In this way, we
automatically and rigorously prove the results for bosonic
channels, following the methods that were originally designed in
Refs.~\cite{QKDpaper,netpaper,longVersion}.

The paper is organized as follows. In Sec.~\ref{GeneralSECTION}\ we present
preliminary notions on adaptive protocols, besides the tools of simulation and
stretching for channels~\cite{QKDpaper} and networks~\cite{netpaper}. The
expert reader can skip this part and directly start from
Sec.~\ref{SECmultipleNETs}\ which provides general description of the various
configurations considered in this work. Secs.~\ref{SecMULTIunicast}
and~\ref{SecMULTIunicast2} consider multiple-unicast settings under single-
and multi-path routing strategies for the quantum systems.
Sec.~\ref{SECmulticastSINGLE} investigates the case of multicast communication
from a sender to multiple receivers. Following Sec.~\ref{SECmulticastMANY}
considers multiple-multicast communication between many senders and many
receivers. Finally, Sec.~\ref{SECconclusions}\ is for conclusions.

\section{Preliminaries\label{GeneralSECTION}}

\subsection{Adaptive point-to-point protocols}

\begin{figure}[pth]
\vspace{-2.3cm}
\par
\begin{center}
\includegraphics[width=0.50\textwidth]{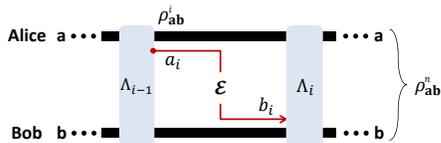} \vspace{-2.9cm}
\end{center}
\caption{Generic $i$th transmission through channel $\mathcal{E}$ in a
point-to-point adaptive protocol. The transmission $a_{i}\rightarrow b_{i}$ is
interleaved by two adaptive LOCCs, $\Lambda_{i-1}$ and $\Lambda_{i}$,
performed by Alice and Bob on their local registers $\mathbf{a}$ and
$\mathbf{b}$.}%
\label{longPIC}%
\end{figure}

Let us first discuss the general structure of an adaptive
point-to-point protocol $\mathcal{P}$ through a quantum channel
$\mathcal{E}$, following the notation from Ref.~\cite{QKDpaper}.
Alice has a local register of quantum systems $\mathbf{a}$ and Bob
has another local register $\mathbf{b}$; these are prepared in a
state $\rho_{\mathbf{ab}}^{0}$ by means of local operations (LOs)
assisted by two-way classical communication (CC), also known as
adaptive LOCCs. After the first adaptive LOCC $\Lambda_{0}$, Alice
selects a system $a_{1}\in\mathbf{a}$ and sends it to Bob through
the quantum channel $\mathcal{E}$. Once Bob receives the output
$b_{1}$, this is included in his register
$b_{1}\mathbf{b}\rightarrow\mathbf{b}$ and another adaptive LOCC
$\Lambda_{1}$ is performed by the parties. The second transmission
starts by selecting another system $a_{2}\in\mathbf{a}$ which is
sent through $\mathcal{E}$ whose output $a_{2}$\ is received by
Bob. Bob updates his register
$b_{2}\mathbf{b}\rightarrow\mathbf{b}$ and another adaptive LOCC
$\Lambda_{2}$ is performed. The generic $i$-th transmission is
shown in Fig.~\ref{longPIC}. After $n$ uses, Alice and Bob have
implemented an adaptive
protocol $\mathcal{P}$ defined by the sequence of LOCCs $\{\Lambda_{0}%
,\Lambda_{1}\ldots\}$ and providing an output state $\rho_{\mathbf{ab}}^{n}$
close in trace norm to a target state $\phi^{n}$ with $nR_{n}^{\varepsilon}$
(target) bits, i.e., such that $\left\Vert \rho_{\mathbf{ab}}^{n}-\phi
^{n}\right\Vert \leq\varepsilon$.

If we now consider the limit for large $n$ (asymptotic rate) and small
$\varepsilon$ (weak converse), and then we optimize over the protocols
$\mathcal{P}$, we define the two-way assisted capacity of $\mathcal{E}$, i.e.,%
\begin{equation}
\mathcal{C}(\mathcal{E}):=\sup_{\mathcal{P}}\lim_{\varepsilon,n}%
R_{n}^{\varepsilon}.
\end{equation}
Assume that the target $\phi^{n}$ is a maximally-entangled state, so that the
target bits are entanglement bits (ebits). In this case, $\mathcal{C}%
(\mathcal{E})$ corresponds to the two-way entanglement-distribution capacity
$D_{2}(\mathcal{E})$, which is also equal to the two-way quantum capacity
$Q_{2}(\mathcal{E})$. Assume instead that $\phi^{n}$ is a private
state~\cite{KD1}, so that the target bits are private bits, then
$\mathcal{C}(\mathcal{E})$ corresponds to the secret key capacity
$K(\mathcal{E})\geq D_{2}(\mathcal{E})$.

\subsection{LOCC simulation of quantum channels}

One may follow the general recipe of Ref.~\cite{QKDpaper} to simplify an
adaptive protocol over $\mathcal{E}$ and write a single-letter upper bound for
the two-way assisted capacity $\mathcal{C}(\mathcal{E})$. The first step is
the simulation of the channel $\mathcal{E}$ by means of an LOCC $\mathcal{T}$
and some resource state $\sigma$. For any channel $\mathcal{E}$, we may always
find a simulation such that
\begin{equation}
\mathcal{E}(\rho)=\mathcal{T}(\rho\otimes\sigma). \label{sigma}%
\end{equation}
A channel simulable with resource state $\sigma$\ may also be called
\textquotedblleft$\sigma$-stretchable\textquotedblright. It is important to
note that the simulation may also be asymptotic, so that we have a sequence of
LOCCs $\mathcal{T}^{\mu}$ and resource states $\sigma^{\mu}$ such that we may
write the point-wise limit~\cite{QKDpaper}%
\begin{equation}
\mathcal{E}(\rho)=\lim_{\mu}\mathcal{T}^{\mu}(\rho\otimes\sigma^{\mu}).
\label{asymptotic}%
\end{equation}

A very convenient simulation holds for those channels\ commuting with the
teleportation correction unitaries, which are (generalized)\ Pauli operators
in finite dimension and phase-space displacements in continuous variable
systems~\cite{telereview}. By definition, a quantum channel $\mathcal{E}$ is
called teleportation-covariant, or just \textquotedblleft
telecovariant\textquotedblright\ when, for any teleportation unitary $U$, we
may write%
\begin{equation}
\mathcal{E}(U\rho U^{\dagger})=V\mathcal{E}(\rho)V^{\dagger}~,
\label{stretchability}%
\end{equation}
for another (generally-different) unitary $V$. Note that Pauli
channels~\cite{NiCh}, erasure channels and bosonic Gaussian
channels~\cite{review} are all telecovariant.

For a telecovariant channel $\mathcal{E}$, we write the simulation
\begin{equation}
\mathcal{E}(\rho)=\mathcal{T}_{\text{tele}}(\rho\otimes\sigma_{\mathcal{E}}),
\label{kkkll}%
\end{equation}
where $\mathcal{T}_{\text{tele}}$\ is a teleportation
protocol~\cite{teleBENNETT,Samtele,Samtele2,telereview} and $\sigma
_{\mathcal{E}}:=\mathcal{I}\otimes\mathcal{E}(\Phi)$ is the Choi matrix of the
channel, with $\Phi$ being a maximally entangled state. For bosonic Gaussian
channels, the Choi matrix is asymptotic, i.e., defined by the $\mu$-limit of
the sequence $\sigma_{\mathcal{E}}^{\mu}:=\mathcal{I}\otimes\mathcal{E}%
(\Phi^{\mu})$, where $\Phi^{\mu}$ is a two-mode squeezed vacuum (TMSV) state
with variance parameter $\mu$~\cite{review}. Thus, we may write the asymptotic
simulation%
\begin{equation}
\mathcal{E}(\rho)=\lim_{\mu}\mathcal{T}_{\text{tele}}^{\mu}(\rho\otimes
\sigma_{\mathcal{E}}^{\mu}), \label{asymptotics}%
\end{equation}
where $\mathcal{T}_{\text{tele}}^{\mu}$ is a sequence of teleportation-LOCCs.

\subsection{Stretching and single-letter bound}

In an adaptive protocol, we may replace each transmission through the channel
$\mathcal{E}$ with its simulation $(\mathcal{T},\sigma)$. Then, as shown in
Ref.~\cite{QKDpaper}, we may collapse all the simulation LOCCs $\mathcal{T}$
and the adaptive LOCCs of the protocol $\{\Lambda_{0},\Lambda_{1}\ldots\}$
into a single trace-preserving LOCC $\bar{\Lambda}$. In this way, the
$n$-use\ output state of the protocol can be decomposed as%
\begin{equation}
\rho_{\mathbf{ab}}^{n}=\bar{\Lambda}\left(  \sigma^{\otimes n}\right)  .
\label{StretchingMAIN}%
\end{equation}
If the simulation of the channel is asymptotic, the stretching takes the form
$\rho_{\mathbf{ab}}^{n}=\lim_{\mu}\bar{\Lambda}_{\mu}(\sigma^{\mu\otimes n})$
for a sequence of trace-preserving LOCC $\bar{\Lambda}_{\mu}$ and resource
states $\sigma^{\mu}$. See Ref.~\cite{QKDpaper,TQC,BK2} for a precise
formulation of this limit where the simulation error is explicitly taken into account.

Suppose that we want to compute an entanglement measure over the output state.
In particular, let us consider the REE. For a quantum state $\rho$, this
is~\cite{RMPrelent}
\begin{equation}
E_{\mathrm{R}}(\rho)=\inf_{\gamma\in\text{\textrm{SEP}}}S(\rho||\gamma),
\label{REEbona}%
\end{equation}
where $\gamma$ is an arbitrary separable state and $S$ is the quantum relative
entropy $S(\rho||\gamma):=\mathrm{Tr}\left[  \rho(\log_{2}\rho-\log_{2}%
\gamma)\right]  $. More weakly, if we consider an asymptotic state
$\sigma:=\lim_{\mu}\sigma^{\mu}$, the previous definition can be extended as%
\begin{equation}
E_{\text{\textrm{R}}}(\sigma)=\underset{\mu\rightarrow+\infty}{\lim\inf
}~E_{\text{\textrm{R}}}(\sigma^{\mu})=\inf_{\gamma^{\mu}}~\underset
{\mu\rightarrow+\infty}{\lim\inf}~S(\sigma^{\mu}||\gamma^{\mu}),
\label{extendedREE}%
\end{equation}
where $\gamma^{\mu}$ is a converging sequence of separable
states~\cite{QKDpaper}.

Because the REE\ is monotonic under trace-preserving LOCCs (data processing)
and sub-additive over tensor products, we may write
\begin{equation}
E_{\mathrm{R}}(\rho_{\mathbf{ab}}^{n})=E_{\mathrm{R}}\left[  \bar{\Lambda
}\left(  \sigma^{\otimes n}\right)  \right]  \leq E_{\mathrm{R}}\left(
\sigma^{\otimes n}\right)  \leq nE_{\mathrm{R}}\left(  \sigma\right)  .
\end{equation}
Now recall that the REE is also asymptotically continuous. This means that for
$\rho_{\mathbf{ab}}^{n}$ and $\phi^{n}$ such that $\left\Vert \rho
_{\mathbf{ab}}^{n}-\phi^{n}\right\Vert \leq\varepsilon$, we may write
\begin{equation}
|E_{\mathrm{R}}(\phi^{n})-E_{\mathrm{R}}(\rho_{\mathbf{ab}}^{n})|\leq
\delta(d,\varepsilon), \label{deltaTERM}%
\end{equation}
where $\delta(\varepsilon,d)$ depends the $\varepsilon$-closeness, and the
dimension $d$ of the total Hilbert space. In the limit of large $n$ and small
$\varepsilon$ (weak converse), we can neglect $\delta(\varepsilon,d)/n$. This
is a straightforward application of the exponential scaling of the dimension
$d$ shown in Refs.~\cite{Matthias1a,Matthias2a} for DV systems and extended to
CV systems in Ref.~\cite{QKDpaper}. For a simplified treatment for CV systems
see also Ref.~\cite{TQC}. Because $E_{\mathrm{R}}(\phi^{n})\geq nR_{n}%
^{\varepsilon}$~\cite{KD1}, we therefore have%
\begin{equation}
R_{n}^{\varepsilon}\leq E_{\mathrm{R}}\left(  \sigma\right)  +n^{-1}%
\delta(d,\varepsilon),
\end{equation}
which leads to the single-letter upper bound~\cite{QKDpaper}
\begin{equation}
\mathcal{C}(\mathcal{E})\leq E_{\mathrm{R}}(\sigma). \label{UB1}%
\end{equation}
In particular, for telecovariant $\mathcal{E}$, we may write
\begin{equation}
\mathcal{C}(\mathcal{E})\leq E_{\mathrm{R}}(\sigma_{\mathcal{E}}), \label{UB2}%
\end{equation}
with a suitable asymptotic formulation for bosonic Gaussian channels based on
Eq.~(\ref{extendedREE}).

Among telecovariant channels, the \textquotedblleft
distillable\textquotedblright\ ones are those for which we may write
$E_{\mathrm{R}}(\sigma_{\mathcal{E}})=D_{1}(\sigma_{\mathcal{E}})$, where
$D_{1}(\sigma_{\mathcal{E}})$ is the distillable entanglement of the
(possibly-asymptotic) channel's Choi matrix $\sigma_{\mathcal{E}}$ via one-way
CCs, forward or backward. This is lower-bounded by both the coherent
\cite{Schu96,Lloyd97} and reverse coherent~\cite{RevCohINFO,ReverseCAP}
information of the Choi matrix. For a distillable channel, the two-way
capacities coincide and are given by%
\begin{equation}
\mathcal{C}(\mathcal{E})=E_{\mathrm{R}}(\sigma_{\mathcal{E}})=D_{1}%
(\sigma_{\mathcal{E}}). \label{coincidence}%
\end{equation}
This is the case for a number of channels, including the dephasing channel,
the erasure channel, the pure-loss channel and the quantum-limited amplifier.
For a pure-loss channel with transmissivity $\eta$, the two-way capacity is
simply given by the PLOB\ bound~\cite[Eq.~(19)]{QKDpaper}%
\begin{equation}
\mathcal{C}(\eta)=-\log_{2}(1-\eta)~. \label{formCloss}%
\end{equation}
As secret-key capacity, this bounds the maximum rate of any point-to-point QKD protocol.

\subsection{Notation for quantum networks}

We model a quantum communication network $\mathcal{N}$ as an undirected finite
graph~\cite{Slepian} $\mathcal{N}=(P,E)$, where $P$ is the set of points or
nodes, and $E$ is the set of undirected edges. Every point can be identified
with a corresponding local register $\mathbf{p}$ of quantum systems. The
existence of an edge $(\mathbf{x},\mathbf{y})$, between two points
$\mathbf{x}$\ and $\mathbf{y}$, means that there is a physical quantum channel
$\mathcal{E}_{\mathbf{xy}}$ between them (which can be forward $\mathcal{E}%
_{\mathbf{x\rightarrow y}}$\ or backward $\mathcal{E}_{\mathbf{y}%
\rightarrow\mathbf{x}}$). For points $\mathbf{p}_{i}$ and $\mathbf{p}_{j}$, we
also use the short-hand notation $\mathcal{E}_{ij}:=\mathcal{E}_{\mathbf{p}%
_{i}\mathbf{p}_{j}}$. We use the notation $\mathbf{a}$ and $\mathbf{b}$ for
Alice and Bob, respectively. A path or route between these two end-points is a
sequence of edges $(\mathbf{a},\mathbf{p}_{i}),\cdots,(\mathbf{p}%
_{j},\mathbf{b})$, that we may simply denote as $\mathbf{a}-\mathbf{p}%
_{i}-\cdots-\mathbf{p}_{j}-\mathbf{b}$. For a route with $N$ middle points (or
repeaters), we have $N+1$ edges and, therefore, a corresponding sequence of
$N+1$ channels $\{\mathcal{E}_{0},\cdots,\mathcal{E}_{k},\cdots,\mathcal{E}%
_{N}\}$ from Alice to Bob.

There are different possible routes between two end-points. For this reason,
they may also use different routing strategies. In single-path routing, Alice
and Bob exploit a single route in each use of the network, and this route can
be changed for the different network uses. In multi-path routing, Alice and
Bob exploit many routes in parallel in each use of the network. In particular,
they may adopt a flooding strategy~\cite{flooding} where each edge of the
network is used exactly once in each end-to-end transmission. In both cases,
we assume that the quantum protocols are adaptive, meaning that each
transmission through each channel is interleaved with a network adaptive
LOCCs, where all points of the network apply LOs on their local registers
assisted by unlimited two-way CC with all the other points of the network.

An entanglement cut $C$ of the quantum network $\mathcal{N}$ is a bipartition
$(\mathbf{A},\mathbf{B})$ of all the points $P$\ such that $\mathbf{a}%
\in\mathbf{A}$ and $\mathbf{b}\in\mathbf{B}$. Correspondingly, the cut-set
$\tilde{C}$ of $C$ is the ensemble of edges across the bipartition, i.e.,
$\tilde{C}=\{(\mathbf{x},\mathbf{y})\in E:\mathbf{x}\in\mathbf{A}%
,\mathbf{y}\in\mathbf{B}\}$. It is clear that $\tilde{C}$\ also identifies an
ensemble of channels $\{\mathcal{E}_{\mathbf{xy}}\}_{(\mathbf{x}%
,\mathbf{y})\in\tilde{C}}$. Given a cut, we may also consider the
complementary sets
\begin{align}
\tilde{A}  &  =\{(\mathbf{x},\mathbf{y})\in E:\mathbf{x,y}\in\mathbf{A}\},\\
\tilde{B}  &  =\{(\mathbf{x},\mathbf{y})\in E:\mathbf{x,y}\in\mathbf{B}\},
\end{align}
so that $\tilde{A}\cup\tilde{B}\cup\tilde{C}=E$.

Given an undirected network $\mathcal{N}=(P,E)$ we can introduce an
orientation by transforming it in a directed graph. One can choose a direction
for all edges so that a generic edge $(\mathbf{x},\mathbf{y})$ becomes an
ordered pair where $\mathbf{x}$ is the \textquotedblleft
tail\textquotedblright\ and $\mathbf{y}$ is the \textquotedblleft
head\textquotedblright. In choosing these directions, we keep Alice as tail
and Bob as head, so that the quantum network can be represented as a flow
network where Alice is the\textit{\ source} and Bob is the\textit{\ sink}%
~\cite{Harris,Ford,ShannonFLOW,Karp,Dinic}. There are $\mathcal{O}(2^{|E|})$
possible orientations. Given an orientation of $\mathcal{N}$, there is a
corresponding flow network $\mathcal{N}_{D}=(P,E_{D})$, where $E_{D}$ is the
set of directed edges. Then, for arbitrary point $\mathbf{p}$, we define its
out-neighborhood as the set of heads going from $\mathbf{p}$%
\begin{equation}
N^{\text{out}}(\mathbf{p})=\{\mathbf{x}\in P:(\mathbf{p},\mathbf{x})\in
E_{D}\},
\end{equation}
and its in-neighborhood as the set of tails going into $\mathbf{p}$%
\begin{equation}
N^{\text{in}}(\mathbf{p})=\{\mathbf{x}\in P:(\mathbf{x},\mathbf{p})\in
E_{D}\}.
\end{equation}
A multi-message quantum multicast from point $\mathbf{p}$ is a
point-to-multipoint connection from $\mathbf{p}$ to part of its
out-neighborhood $N^{\text{out}}(\mathbf{p})$, so that different messages
(quantum states or keys) are received by the receiving points. It is a
single-message multicast if the messages coincide.

\subsection{Simulation and stretching of a network}

\begin{figure}[ptbh]
\vspace{-1.3cm}
\par
\begin{center}
\includegraphics[width=0.5\textwidth] {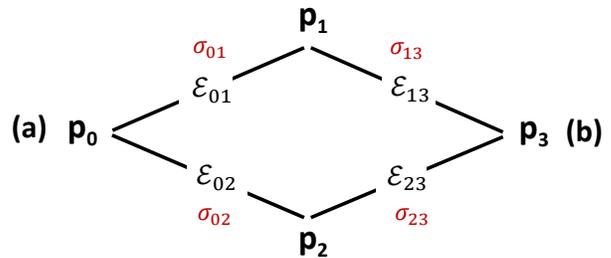} \vspace{-2.2cm}
\end{center}
\caption{Network simulation. Consider a simple four-point quantum network
$\mathcal{N}=(\{\mathbf{p}_{0},\mathbf{p}_{1},\mathbf{p}_{2},\mathbf{p}%
_{3}\},E)$ with end points $\mathbf{p}_{0}=\mathbf{a}$ (Alice) $\mathbf{p}%
_{3}=\mathbf{b}$ (Bob). Edge $(\mathbf{p}_{i},\mathbf{p}_{j})$\ has an
associated quantum channel $\mathcal{E}_{ij}$. By simulating each channel
$\mathcal{E}_{ij}$ with a corresponding resource state $\sigma_{ij}$, we
define a resource representation of the network $\sigma(\mathcal{N}%
)=\{\sigma_{01},\sigma_{02},\sigma_{13},\sigma_{23}\}$.}%
\label{diamond}%
\end{figure}

Given an arbitrary network $\mathcal{N}$, we may replace it with its
simulation~\cite{netpaper,longVersion}. In fact, for any edge $(\mathbf{x}%
,\mathbf{y})$, we may replace the quantum channel $\mathcal{E}_{\mathbf{xy}}$
with its simulation $S_{\mathbf{xy}}=(\mathcal{T}_{\mathbf{xy}},\sigma
_{\mathbf{xy}})$ for some LOCC $\mathcal{T}_{\mathbf{xy}}$ and resource state
$\sigma_{\mathbf{xy}}$. Repeating this process for all the edges defines an
LOCC\ simulation of the network $S(\mathcal{N})=\{S_{\mathbf{xy}%
}\}_{(\mathbf{x},\mathbf{y})\in E}$ where all channels $\mathcal{E}%
_{\mathbf{xy}}$ are replaced by resource states $\sigma_{\mathbf{xy}}$. There
is a corresponding resource representation of the network $\sigma
(\mathcal{N})=\{\sigma_{\mathbf{xy}}\}_{(\mathbf{x},\mathbf{y})\in E}$. See
also Fig.~\ref{diamond} for a simple example. In particular, for a
telecovariant network, where all channels are telecovariant, then the
simulation involves teleportation LOCCs and the network can be replaced by its
Choi representation $\sigma(\mathcal{N})=\{\sigma_{\mathcal{E}_{\mathbf{xy}}%
}\}_{(\mathbf{x},\mathbf{y})\in E}$. Here each channel $\mathcal{E}%
_{\mathbf{xy}}$ is replaced by its (possibly-asymptotic) Choi matrix
$\sigma_{\mathcal{E}_{\mathbf{xy}}}$. A quantum network is said to be
distillable if it is connected by distillable channels.

Given an arbitrary adaptive protocol implemented over the quantum network
$\mathcal{N}$, we can use the network simulation $\sigma(\mathcal{N})$ to
stretch the protocol and decompose the total output state $\rho
_{\mathbf{a\ldots b}}^{n}$ of network after $n$ uses as follows%
\begin{equation}
\rho_{\mathbf{a\ldots b}}^{n}=\bar{\Lambda}\left[  \underset{(\mathbf{x}%
,\mathbf{y})\in E}{%
{\textstyle\bigotimes}
}~\sigma_{\mathbf{xy}}^{\otimes n_{\mathbf{xy}}}\right]  ,
\label{LemmaNETstretching}%
\end{equation}
where $\bar{\Lambda}$\ is a trace-preserving LOCC and $n_{\mathbf{xy}}$ is the
number of uses of the edge $(\mathbf{x},\mathbf{y})$. In particular, we have
$n_{\mathbf{xy}}\leq n$ ($n_{\mathbf{xy}}=n$) for a single-path (flooding)
protocol. In other words, introducing the probabilities $p_{\mathbf{xy}%
}:=n_{\mathbf{xy}}/n$ we have $p_{\mathbf{xy}}\leq1$ ($=1$) for a single-path
(flooding) protocol~\cite{netpaper,longVersion}.

Tracing out all the network points except the two end-points, from
Eq.~(\ref{LemmaNETstretching}) we get Alice and Bob's shared state
$\rho_{\mathbf{ab}}^{n}$. For any entanglement cut $C$ and corresponding
cut-set $\tilde{C}$, we may write a better decomposition for Alice and Bob's
output state. This is given by%
\begin{equation}
\rho_{\mathbf{ab}}^{n}(C)=\bar{\Lambda}_{\mathbf{ab}}\left[  \underset
{(\mathbf{x},\mathbf{y})\in\tilde{C}}{%
{\textstyle\bigotimes}
}~\sigma_{\mathbf{xy}}^{\otimes n_{\mathbf{xy}}}\right]  , \label{cutEQ}%
\end{equation}
where $\bar{\Lambda}_{\mathbf{ab}}$ is a trace-preserving LOCC with respect to
Alice and Bob. Previous Eqs.~(\ref{LemmaNETstretching}) and~(\ref{cutEQ}) can
be extended to asymptotic resource states by introducing suitable limits. See
Ref.~\cite{netpaper,longVersion}\ for more details on these methods.

\section{Multiple senders and receivers\label{SECmultipleNETs}}

One of the basic working mechanisms in a quantum communication network is the
unicast setting, based on a single sender $\mathbf{a}$ and a single receiver
$\mathbf{b}$. However, in general, we may consider multiple senders
$\{\mathbf{a}_{i}\}$ and receivers $\{\mathbf{b}_{j}\}$, which may
simultaneously communicate according to various configurations. For
simplicity, these sets are intended to be disjoint $\{\mathbf{a}_{i}%
\}\cap\{\mathbf{b}_{j}\}=\emptyset$, so that an end-point cannot be sender and
receiver at the same time. It is clear that all the results from
Ref.~\cite{netpaper,longVersion}, derived for the two basic routing
strategies, provide general upper bounds which are still valid for the
individual end-to-end capacities associated with each sender-receiver pair
$(\mathbf{a}_{i},\mathbf{b}_{i})$ in the various settings with multiple end-points.

In the following sections, we start with the multiple-unicast quantum network.
This consists of $M$ Alices $\{\mathbf{a}_{1},\ldots,\mathbf{a}_{M}\}$ and $M$
Bobs $\{\mathbf{b}_{1},\ldots,\mathbf{b}_{M}\}$, with the generic $i$th Alice
$\mathbf{a}_{i}$ communicating with a corresponding $i$th Bob $\mathbf{b}_{i}%
$. This case can be studied by assuming single-path routing
(Sec.~\ref{SecMULTIunicast}) or multipath routing (Sec.~\ref{SecMULTIunicast2}%
). Besides the general bounds inherited from the unicast scenario, we derive a
specific set of upper bounds for the rates that are simultaneously achievable
by all parties.

Another important case is the multicast (multi-message) quantum network, where
a single sender simultaneously communicates with $M\geq1$ receivers, e.g., for
distributing $M$ different states or keys. By its nature, this is studied
under multipath routing (Sec.~\ref{SECmulticastSINGLE}). In this setting, an
interesting variant is the distribution of the same key to all receivers
(single-message multicast).

\begin{figure}[ptbh]
\vspace{-1.0cm}
\par
\begin{center}
\includegraphics[width=0.37\textwidth] {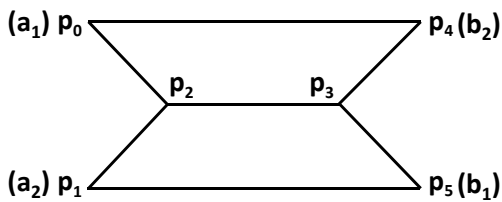} \vspace{-1.3cm}
\end{center}
\caption{Butterfly quantum network. (i)~An example of multiple-unicast is
considering two sender-receiver pairs, e.g., Alice $\mathbf{a}_{1}$
communicating with Bob $\mathbf{b}_{1}$, and Alice $\mathbf{a}_{2}$ with Bob
$\mathbf{b}_{2}$. Single-path routing corresponds to the simultaneous use of
two end-to-end routes, e.g., $(\mathbf{a}_{1})\mathbf{p}_{0}-\mathbf{p}%
_{2}-\mathbf{p}_{3}-\mathbf{p}_{5}(\mathbf{b}_{1})$ and $(\mathbf{a}%
_{2})\mathbf{p}_{1}-\mathbf{p}_{2}-\mathbf{p}_{3}-\mathbf{p}_{4}%
(\mathbf{b}_{2})$. Multipath routing corresponds to choosing a network
orientation, where the end-points may also act as relays. Each point of the
network multicasts multiple messages to its out-neighborhood. For instance, we
may have the point-to-multipoint multicasts: $\mathbf{p}_{0}\rightarrow
\{\mathbf{p}_{2},\mathbf{p}_{4}\}$, $\mathbf{p}_{1}\rightarrow\{\mathbf{p}%
_{2},\mathbf{p}_{5}\}$, $\mathbf{p}_{2}\rightarrow\mathbf{p}_{3}$, and
$\mathbf{p}_{3}\rightarrow\{\mathbf{p}_{4},\mathbf{p}_{5}\}$. (ii)~An example
of network multicast is Alice $\mathbf{a}_{1}$ communicating with the two Bobs
$\{\mathbf{b}_{1},\mathbf{b}_{2}\}$ via multipath routing. In general, the
messages (states, keys) can be different. (iii)~In a multiple-multicast, Alice
$\mathbf{a}_{1}$ communicates with $\{\mathbf{b}_{1},\mathbf{b}_{2}\}$, and
Alice $\mathbf{a}_{2}$ communicates with the same destination set
$\{\mathbf{b}_{1},\mathbf{b}_{2}\}$. In general, the messages (states, keys)
can be different.}%
\label{butterfly}%
\end{figure}

More generally, we may consider a multiple-multicast (multi-message) quantum
network. Here we have $M_{A}\geq1$ senders and $M_{B}\geq1$ receivers, and
each sender communicates simultaneously with the entire set of receivers
communicating different states or keys~(Sec.~\ref{SECmulticastMANY}). In a
private communication scenario, this corresponds to the distribution of
$M_{A}M_{B}$ different keys. For a description of these configurations, see
the simple example of the butterfly quantum network in Fig.~\ref{butterfly}.

\section{Multiple-unicast quantum networks with single-path
routing\label{SecMULTIunicast}}

Let us start by considering two Alice-Bob pairs $(\mathbf{a}_{1}%
,\mathbf{b}_{1})$ and $(\mathbf{a}_{2},\mathbf{b}_{2})$, since the extension
to arbitrary number of pairs is immediate. We may easily formulate network
protocols which are based on single-path routing. In this case, each
sequential use of the network involves the transmission of quantum systems
along two (potentially-overlapping) routes%
\begin{equation}
\omega_{1}:\mathbf{a}_{1}-\cdots-\mathbf{b}_{1},~~~\omega_{2}:\mathbf{a}%
_{2}-\cdots-\mathbf{b}_{2},
\end{equation}
where each transmission through an edge is assisted by network LOCCs. The
routes are updated use after use.

After $n$ uses, the output of the double-unicast network protocol
$\mathcal{P}_{\text{2-unicast}}$ is a state $\rho_{\mathbf{a}_{1}%
\mathbf{a}_{2}\mathbf{b}_{1}\mathbf{b}_{2}}^{n}$ which is $\varepsilon$-close
in trace norm to a target state
\begin{equation}
\phi:=\phi_{\mathbf{a}_{1}\mathbf{b}_{1}}^{\otimes nR_{1}^{\varepsilon,n}%
}\otimes\phi_{\mathbf{a}_{2}\mathbf{b}_{2}}^{\otimes nR_{2}^{\varepsilon,n}},
\label{targettt}%
\end{equation}
where $\phi_{\mathbf{a}_{i}\mathbf{b}_{i}}$ is a one-bit state (private bit or
ebit) for the pair $(\mathbf{a}_{i},\mathbf{b}_{i})$ and $nR_{i}%
^{\varepsilon,n}$ the number of its copies. Taking the limit of large $n$,
small $\varepsilon$ (weak converse) and optimizing over all protocols
$\mathcal{P}_{\text{2-unicast}}$, we define the capacity region as the closure
of the set of the achievable asymptotic rates $(R_{1},R_{2})$. In general, for
$M$ sender-receiver pairs, we have an $M$-tuple of achievable rates
$(R_{1},\ldots,R_{M})$. Depending on the task of the protocol (i.e., the
target state), these rates refer to end-to-end entanglement distillation
(equivalently, error-free quantum communication) or secret-key generation.

Before proceeding, let us first introduce more general types of entanglement
cuts of the quantum network. Given two sets of senders $\{\mathbf{a}_{i}\}$
and receivers $\{\mathbf{b}_{i}\}$, we adopt the notation $C:\{\mathbf{a}%
_{i}\}|\{\mathbf{b}_{i}\}$ for a cut $C=(\mathbf{A},\mathbf{B})$ such that
$\{\mathbf{a}_{i}\}\subset\mathbf{A}$ and $\{\mathbf{b}_{i}\}\subset
\mathbf{B}$. Similarly, we write $C:\mathbf{a}_{i}|\mathbf{b}_{i}$ for a cut
with $\mathbf{a}_{i}\in\mathbf{A}$ and $\mathbf{b}_{i}\in\mathbf{B}$, and
$C:\mathbf{a}_{i}\mathbf{a}_{j}|\mathbf{b}_{i}\mathbf{b}_{j}$ for a cut with
$\{\mathbf{a}_{i},\mathbf{a}_{j}\}\subset\mathbf{A}$ and $\{\mathbf{b}%
_{i},\mathbf{b}_{j}\}\subset\mathbf{B}$. Define also the single-edge flow of
entanglement (REE) trough a cut as%
\begin{equation}
E_{\mathrm{R}}(C):=\max_{(\mathbf{x},\mathbf{y})\in\tilde{C}}E_{R}%
(\sigma_{\mathbf{xy}}),
\end{equation}
where $\sigma_{\mathbf{xy}}$ is a resource state associated with an edge
$(\mathbf{x},\mathbf{y})$ in the cut-set $\tilde{C}$, under some simulation of
the network. We can then state the following result.

\begin{theorem}
[Multi-unicast with single paths]\label{TheomultipleUNICAST}Let us consider a
multiple-unicast quantum network $\mathcal{N}=(P,E)$ with $M$ sender-receiver
pairs $(\mathbf{a}_{i},\mathbf{b}_{i})$ communicating by means of single-path
routing. Adopt a simulation of the network with a resource representation
$\sigma(\mathcal{N})=\{\sigma_{\mathbf{xy}}\}_{(\mathbf{x},\mathbf{y})\in E}$.
In particular, $\sigma(\mathcal{N})$\ can be a Choi-representation for a
teleportation-covariant $\mathcal{N}$. We have the following outer bounds for
the capacity region%
\begin{align}
R_{i}  &  \leq\min_{C:\mathbf{a}_{i}|\mathbf{b}_{i}}E_{\mathrm{R}%
}(C)~~\text{for any }i,\label{unicvvv}\\
R_{i}+R_{j}  &  \leq\min_{C:\mathbf{a}_{i}\mathbf{a}_{j}|\mathbf{b}%
_{i}\mathbf{b}_{j}}E_{\mathrm{R}}(C)~~\text{for any }i\neq j
\label{doublehhhh}\\
&  \vdots\nonumber\\
\sum\limits_{i=1}^{M}R_{i}  &  \leq\min_{C:\{\mathbf{a}_{i}\}|\{\mathbf{b}%
_{i}\}}E_{\mathrm{R}}(C),
\end{align}
where $E_{\mathrm{R}}(C)$\ is the single-edge flow of REE through cut $C$. It
is understood that formulations may be asymptotic for quantum networks with
bosonic channels.
\end{theorem}

\textbf{Proof.}~~For simplicity, first consider the case $M=2$, since the
generalization to arbitrary $M$ is straightforward. Let us also consider key
generation, since it automatically provides an upper bound for all the other
tasks. Considering the bipartition $\mathbf{a}_{1}\mathbf{a}_{2}%
|\mathbf{b}_{1}\mathbf{b}_{2}$, the distillable key of the target state $\phi$
in Eq.~(\ref{targettt}) is equal to%
\begin{equation}
K_{\mathrm{D}}(\mathbf{a}_{1}\mathbf{a}_{2}|\mathbf{b}_{1}\mathbf{b}%
_{2})_{\phi}=n(R_{1}^{\varepsilon,n}+R_{2}^{\varepsilon,n}).
\end{equation}
Using the REE with respect to the same bipartition, we may write the upper
bound%
\begin{align}
n(R_{1}^{\varepsilon,n}+R_{2}^{\varepsilon,n})  &  \leq E_{\mathrm{R}%
}(\mathbf{a}_{1}\mathbf{a}_{2}|\mathbf{b}_{1}\mathbf{b}_{2})_{\phi}\nonumber\\
&  \leq E_{\mathrm{R}}(\mathbf{a}_{1}\mathbf{a}_{2}|\mathbf{b}_{1}%
\mathbf{b}_{2})_{\rho^{n}}+\delta(\varepsilon,d),
\end{align}
where the latter inequality comes from the fact that $\rho^{n}:=\rho
_{\mathbf{a}_{1}\mathbf{a}_{2}\mathbf{b}_{1}\mathbf{b}_{2}}^{n}$ is
$\varepsilon$-close to $\phi$. The extra term $\delta(\varepsilon,d)$ depends
on the $\varepsilon$-closeness and the dimension $d$ of the total Hilbert
space, as already discussed in relation to Eq.~(\ref{deltaTERM}). The term
$n^{-1}\delta(d,\varepsilon)$ goes to zero for large $n$ and small
$\varepsilon$. As a result we may write
\begin{equation}
\lim_{\varepsilon,n}(R_{1}^{\varepsilon,n}+R_{2}^{\varepsilon,n})\leq
\underset{n\rightarrow+\infty}{\lim}n^{-1}E_{\mathrm{R}}(\mathbf{a}%
_{1}\mathbf{a}_{2}|\mathbf{b}_{1}\mathbf{b}_{2})_{\rho^{n}}~. \label{todecppp}%
\end{equation}

By simulating and stretching the network, we may write the following
decomposition of the output state%
\begin{equation}
\rho_{\mathbf{a}_{1}\mathbf{a}_{2}\mathbf{b}_{1}\mathbf{b}_{2}}^{n}%
=\bar{\Lambda}_{\mathbf{a}_{1}\mathbf{a}_{2}\mathbf{b}_{1}\mathbf{b}_{2}%
}\left[  \underset{(\mathbf{x},\mathbf{y})\in E}{%
{\textstyle\bigotimes}
}~\sigma_{\mathbf{xy}}^{\otimes n_{\mathbf{xy}}}\right]  , \label{cmtoo}%
\end{equation}
where $n_{\mathbf{xy}}=np_{\mathbf{xy}}$ is the number of uses of edge
$(\mathbf{x},\mathbf{y})$ and $\bar{\Lambda}_{\mathbf{a}_{1}\mathbf{a}%
_{2}\mathbf{b}_{1}\mathbf{b}_{2}}$ is a trace-preserving LOCC, which is local
with respect to the bipartition $\mathbf{a}_{1}\mathbf{a}_{2}|\mathbf{b}%
_{1}\mathbf{b}_{2}$. By inserting entanglement cuts which disconnect the
senders and receivers, we reduce the number of resource states appearing in
Eq.~(\ref{cmtoo}) while preserving the locality of the LOCC with respect to
the bipartition of the end-points. In other words, for any cut $C:\mathbf{a}%
_{1}\mathbf{a}_{2}|\mathbf{b}_{1}\mathbf{b}_{2}$ we may write%
\begin{equation}
\rho_{\mathbf{a}_{1}\mathbf{a}_{2}\mathbf{b}_{1}\mathbf{b}_{2}}^{n}%
(C)=\bar{\Lambda}_{\mathbf{a}_{1}\mathbf{a}_{2}\mathbf{b}_{1}\mathbf{b}_{2}%
}^{C}\left[  \underset{(\mathbf{x},\mathbf{y})\in\tilde{C}}{%
{\textstyle\bigotimes}
}~\sigma_{\mathbf{xy}}^{\otimes n_{\mathbf{xy}}}\right]  .
\end{equation}

Using the latter decomposition in Eq.~(\ref{todecppp}), we obtain%
\begin{align}
\lim_{\varepsilon,n}(R_{1}^{\varepsilon,n}+R_{2}^{\varepsilon,n})  &
\leq\underset{n\rightarrow+\infty}{\lim}~n^{-1}E_{\mathrm{R}}(\mathbf{a}%
_{1}\mathbf{a}_{2}|\mathbf{b}_{1}\mathbf{b}_{2})_{\rho^{n}(C)}\nonumber\\
&  \leq\underset{n\rightarrow+\infty}{\lim}~n^{-1}\sum\limits_{(\mathbf{x}%
,\mathbf{y})\in\tilde{C}}n_{\mathbf{xy}}E_{\mathrm{R}}(\sigma_{\mathbf{xy}%
})\nonumber\\
&  =\sum\limits_{(\mathbf{x},\mathbf{y})\in\tilde{C}}p_{\mathbf{xy}%
}E_{\mathrm{R}}(\sigma_{\mathbf{xy}})\nonumber\\
&  \leq\max_{(\mathbf{x},\mathbf{y})\in\tilde{C}}E_{\mathrm{R}}(\sigma
_{\mathbf{xy}}):=E_{\mathrm{R}}(C).
\end{align}

By minimizing over the cuts, we derive%
\begin{equation}
\lim_{\varepsilon,n}(R_{1}^{\varepsilon,n}+R_{2}^{\varepsilon,n})\leq
\min_{C:\mathbf{a}_{1}\mathbf{a}_{2}|\mathbf{b}_{1}\mathbf{b}_{2}%
}E_{\mathrm{R}}(C). \label{outerCCC}%
\end{equation}
It is important to note that this bound holds for any protocol $\mathcal{P}%
_{\text{2-unicast}}$, whose details are all collapsed in the LOCC
$\bar{\Lambda}_{\mathbf{a}_{1}\mathbf{a}_{2}\mathbf{b}_{1}\mathbf{b}_{2}}$ and
therefore discarded. Thus, the same bound applies if we optimize over all
protocols, which means that Eq.~(\ref{outerCCC}) provides the following outer
bound for the capacity region%
\begin{align}
R_{1}+R_{2}  &  =\sup_{\mathcal{P}_{\text{2-unicast}}}\lim_{\varepsilon
,n}(R_{1}^{\varepsilon,n}+R_{2}^{\varepsilon,n})\nonumber\\
&  \leq\min_{C:\mathbf{a}_{1}\mathbf{a}_{2}|\mathbf{b}_{1}\mathbf{b}_{2}%
}E_{\mathrm{R}}(C).
\end{align}

Note that, besides this bound, we also have the following unicast bounds for
the individual rates
\begin{equation}
R_{1}\leq\min_{C:\mathbf{a}_{1}|\mathbf{b}_{1}}E_{\mathrm{R}}(C),~~R_{2}%
\leq\min_{C:\mathbf{a}_{2}|\mathbf{b}_{2}}E_{\mathrm{R}}(C).
\end{equation}
These follows directly from the results of Ref.~\cite{netpaper,longVersion} on
the converse bounds for unicast quantum networks. Equivalently, we may
re-derive these bounds here, by setting $R_{2}=0$ or $R_{1}=0$ in the target
state of Eq.~(\ref{targettt}) and repeating the previous derivation. For
instance, for $R_{2}=0$, we have $\phi:=\phi_{\mathbf{a}_{1}\mathbf{b}_{1}%
}^{\otimes nR_{1}^{\varepsilon,n}}\otimes\sigma_{\mathbf{a}_{2}\mathbf{b}_{2}%
}$, where $\sigma_{\mathbf{a}_{2}\mathbf{b}_{2}}$ does not contain target bits
and may be taken to be separable. Therefore, we start from $K_{\mathrm{D}%
}(\mathbf{a}_{1}|\mathbf{b}_{1})_{\phi}=nR_{1}^{\varepsilon,n}$ and repeat the
derivation with respect to $\mathbf{a}_{1}|\mathbf{b}_{1}$.

It is easy to generalize from $M=2$ to arbitrary $M$. For any integer $M$, we
have the target state%
\begin{equation}
\phi:=%
{\textstyle\bigotimes_{i=1}^{M}}
\phi_{\mathbf{a}_{i}\mathbf{b}_{i}}^{\otimes nR_{i}^{\varepsilon,n}}.
\end{equation}
Considering the bipartition $\{\mathbf{a}_{i}\}|\{\mathbf{b}_{i}\}$ and the
corresponding cuts of the network leads to
\begin{equation}
\sum\limits_{i=1}^{M}R_{i}\leq\min_{C:\{\mathbf{a}_{i}\}|\{\mathbf{b}_{i}%
\}}E_{\mathrm{R}}(C),
\end{equation}
where we note that increasing the number of rates reduces the number of
possible cuts in the minimization. To get all the remaining inequalities of
the theorem, we just need to set some of the rates to zero. For instance, for
$R_{i}\neq0$ and $R_{j\neq i}=0$, we get the unicast bounds of
Eq.~(\ref{unicvvv}). For $R_{i}\neq0$, $R_{j\neq i}\neq0$ and $R_{k\neq
i,j}=0$ we get the double-unicast bounds of Eq.~(\ref{doublehhhh}), and so on.
The extension to asymptotic simulations of bosonic channels is achieved via
the weaker definition of REE in Eq.~(\ref{extendedREE}).~$\blacksquare$

Once we have Theorem~\ref{TheomultipleUNICAST}, it is immediate to specify the
results for the case of multiple-unicast distillable networks, for which we
may write $E_{\mathrm{R}}(\sigma_{\mathbf{xy}})=E_{\mathrm{R}}(\sigma
_{\mathcal{E}_{\mathbf{xy}}})=\mathcal{C}_{\mathbf{xy}}$ for each edge
$(\mathbf{x,y})\in E$, where $\mathcal{C}_{\mathbf{xy}}$ is the two-way
capacity of the associated quantum channel $\mathcal{E}_{\mathbf{xy}}$. In
this case, we may directly write
\begin{equation}
E_{\mathrm{R}}(C)=\mathcal{C}(C):=\max_{(\mathbf{x},\mathbf{y})\in\tilde{C}%
}\mathcal{C}_{\mathbf{xy}},
\end{equation}
where $\mathcal{C}(C)$ is the single-edge capacity of cut $C$. Thus, we can
express the bounds of Theorem~\ref{TheomultipleUNICAST} in terms of the
capacities of the cuts, immediately proving the following.

\begin{corollary}
Consider a multiple-unicast quantum network $\mathcal{N}$\ with $M$
sender-receiver pairs $(\mathbf{a}_{i},\mathbf{b}_{i})$ communicating by means
of single-path routing. If the network is distillable, then we may write the
following outer bounds for the capacity region%
\begin{align}
R_{i}  &  \leq\min_{C:\mathbf{a}_{i}|\mathbf{b}_{i}}\mathcal{C}(C)~~\text{for
any }i,\label{distooo}\\
R_{i}+R_{j}  &  \leq\min_{C:\mathbf{a}_{i}\mathbf{a}_{j}|\mathbf{b}%
_{i}\mathbf{b}_{j}}\mathcal{C}(C)~~\text{for any }i\neq j\\
&  \vdots\nonumber\\
\sum\limits_{i=1}^{M}R_{i}  &  \leq\min_{C:\{\mathbf{a}_{i}\}|\{\mathbf{b}%
_{i}\}}\mathcal{C}(C), \label{distoooo}%
\end{align}
where $\mathcal{C}(C)$ is the single-edge capacity of cut $C$.
\end{corollary}

Note that we cannot establish the achievability of the outer bounds in
Eqs.~(\ref{distooo})-(\ref{distoooo}), apart from the case $M=1$. This case in
fact corresponds to a unicast distillable network for which the bound is
achievable by solving the widest path problem~\cite{netpaper,longVersion}. In
general, for $M>1$, achievable lower bounds can be established by combining
the point-to-point composition strategies with classical routing algorithms
that solve the multiple-version of the widest path problem.

\section{Multiple-unicast quantum networks with multipath
routing\label{SecMULTIunicast2}}

Here we consider a quantum network where $M$\ senders $\{\mathbf{a}_{i}\}$ and
$M$ receivers $\{\mathbf{b}_{i}\}$ communicate in a pairwise fashion
$(\mathbf{a}_{i},\mathbf{b}_{i})$ by means of multipath routing. In a
multipath protocol, the points first agree an orientation for the quantum
network. For multiple-unicasts note that both the senders and receivers may
assists one with each other as relays of the network. This means that
$\{\mathbf{a}_{i}\}$ are not necessarily sources and $\{\mathbf{b}_{i}\}$ are
not necessarily sinks, i.e., these sets may have both incoming and outgoing
edges. Given an orientation, each point multicasts to its out-neighborhood
with the assistance of network LOCCs. This flooding process ends when each
edge of the network has been exploited. For the next use, the points may agree
a different orientation, and so on.

The sequence of the orientations together with the sequence of all network
LOCCs (exploited in each orientation) define a multiple-unicast flooding
protocol $\mathcal{P}_{\text{M-unicast}}^{\text{flood}}$. Its output will be a
shared state $\rho_{\{\mathbf{a}_{i}\}\{\mathbf{b}_{i}\}}^{n}$ which is
$\varepsilon$-close to a target state
\begin{equation}
\phi:=%
{\textstyle\bigotimes_{i=1}^{M}}
\phi_{\mathbf{a}_{i}\mathbf{b}_{i}}^{\otimes nR_{i}^{\varepsilon,n}},
\end{equation}
where $\phi_{\mathbf{a}_{i}\mathbf{b}_{i}}$ is a one-bit state (private bit or
ebit) for the pair $(\mathbf{a}_{i},\mathbf{b}_{i})$ and $nR_{i}%
^{\varepsilon,n}$ the number of its copies. By taking the limit of large $n$,
small $\varepsilon$, and optimizing over $\mathcal{P}_{\text{M-unicast}%
}^{\text{flood}}$, we define the capacity region associated with the
achievable rates $(R_{1}^{\text{m}},\ldots,R_{M}^{\text{m}})$ for the various
quantum tasks. We can then state the following result.

\begin{theorem}
[Multi-unicast with multipaths]\label{TheomultipleUNICAST2}Let us consider a
multiple-unicast quantum network $\mathcal{N}=(P,E)$\ with $M$ sender-receiver
pairs $(\mathbf{a}_{i},\mathbf{b}_{i})$ communicating via multipath routing.
Adopt a simulation of the network with a resource representation
$\sigma(\mathcal{N})=\{\sigma_{\mathbf{xy}}\}_{(\mathbf{x},\mathbf{y})\in E}$.
In particular, $\sigma(\mathcal{N})$\ can be a Choi-representation for a
teleportation-covariant $\mathcal{N}$. We have the following outer bounds for
the capacity region%
\begin{align}
R_{i}^{\text{m}}  &  \leq\min_{C:\mathbf{a}_{i}|\mathbf{b}_{i}}E_{\mathrm{R}%
}^{\text{m}}(C)~~\text{for any }i,\label{rtg}\\
R_{i}^{\text{m}}+R_{j}^{\text{m}}  &  \leq\min_{C:\mathbf{a}_{i}\mathbf{a}%
_{j}|\mathbf{b}_{i}\mathbf{b}_{j}}E_{\mathrm{R}}^{\text{m}}(C)~~\text{for any
}i\neq j\label{rtg2}\\
&  \vdots\nonumber\\
\sum\limits_{i=1}^{M}R_{i}^{\text{m}}  &  \leq\min_{C:\{\mathbf{a}%
_{i}\}|\{\mathbf{b}_{i}\}}E_{\mathrm{R}}^{\text{m}}(C), \label{rtg3}%
\end{align}
where $E_{\mathrm{R}}^{\text{m}}(C):=\sum_{(\mathbf{x},\mathbf{y})\in\tilde
{C}}E_{\mathrm{R}}(\sigma_{\mathbf{xy}})$ is the multi-edge flow of REE across
cut $C$. It is understood that formulations may be asymptotic for quantum
networks with bosonic channels.
\end{theorem}

\textbf{Proof.}~~The proof follows the main steps of the one of
Theorem~\ref{TheomultipleUNICAST}. As before, consider key generation. For the
bipartition $\{\mathbf{a}_{i}\}|\{\mathbf{b}_{i}\}$, the distillable key of
the target state $\phi$ is equal to%
\begin{align}
K_{\mathrm{D}}(\{\mathbf{a}_{i}\}|\{\mathbf{b}_{i}\})_{\phi}  &
=n\sum\limits_{i=1}^{M}R_{i}^{\varepsilon,n}\\
&  \leq E_{R}(\{\mathbf{a}_{i}\}|\{\mathbf{b}_{i}\})_{\phi}\\
&  \leq E_{R}(\{\mathbf{a}_{i}\}|\{\mathbf{b}_{i}\})_{\rho^{n}}+\delta
(\varepsilon,d),
\end{align}
which leads to the inequality%
\begin{equation}
\lim_{\varepsilon,n}\sum\limits_{i=1}^{M}R_{i}^{\varepsilon,n}\leq
\underset{n\rightarrow+\infty}{\lim~}n^{-1}E_{R}(\{\mathbf{a}_{i}%
\}|\{\mathbf{b}_{i}\})_{\rho^{n}}~. \label{repkkkj}%
\end{equation}

For any cut $C:\{\mathbf{a}_{i}\}|\{\mathbf{b}_{i}\}$ of the (simulated)
network, we may write the following decomposition of the output state%
\begin{equation}
\rho_{\{\mathbf{a}_{i}\}\{\mathbf{b}_{i}\}}^{n}(C)=\bar{\Lambda}%
_{\{\mathbf{a}_{i}\}\{\mathbf{b}_{i}\}}^{C}\left[  \underset{(\mathbf{x}%
,\mathbf{y})\in\tilde{C}}{%
{\textstyle\bigotimes}
}~\sigma_{\mathbf{xy}}^{\otimes n}\right]  ,
\end{equation}
for some trace-preserving LOCC $\bar{\Lambda}_{\{\mathbf{a}_{i}\}\{\mathbf{b}%
_{i}\}}^{C}$. Note that here we have $n_{\mathbf{xy}}=n$.\ By replacing
$\rho^{n}=\rho_{\{\mathbf{a}_{i}\}\{\mathbf{b}_{i}\}}^{n}(C)$ in
Eq.~(\ref{repkkkj}), we therefore get%
\begin{equation}
\lim_{\varepsilon,n}\sum\limits_{i=1}^{M}R_{i}^{\varepsilon,n}\leq
\sum\limits_{(\mathbf{x},\mathbf{y})\in\tilde{C}}E_{\mathrm{R}}(\sigma
_{\mathbf{xy}}):=E_{\mathrm{R}}^{\text{m}}(C).
\end{equation}

The next step is to minimize over the cuts, leading to%
\begin{equation}
\lim_{\varepsilon,n}\sum\limits_{i=1}^{M}R_{i}^{\varepsilon,n}\leq
\min_{C:\{\mathbf{a}_{i}\}|\{\mathbf{b}_{i}\}}E_{\mathrm{R}}^{\text{m}}(C).
\end{equation}
Since the latter inequality holds for any protocol $\mathcal{P}%
_{\text{M-unicast}}^{\text{flood}}$, it can be extended to the achievable
rates
\begin{align}
\sum\limits_{i=1}^{M}R_{i}^{\text{m}}  &  =\sup_{\mathcal{P}_{\text{M-unicast}%
}^{\text{flood}}}\lim_{\varepsilon,n}\sum\limits_{i=1}^{M}R_{i}^{\varepsilon
,n}\nonumber\\
&  \leq\min_{C:\{\mathbf{a}_{i}\}|\{\mathbf{b}_{i}\}}E_{\mathrm{R}}^{\text{m}%
}(C).
\end{align}

Finally, by setting some of the rates equal to zero in the target state, we
may repeat the procedure with respect to different bipartitions and derive all
the remaining conditions in Eqs.~(\ref{rtg})-(\ref{rtg3}). The extension to
asymptotic simulations of bosonic channels is achieved by adopting the weaker
definition of the REE.~$\blacksquare$

It is immediate to specify the result for distillable networks for which we
may directly write
\begin{equation}
E_{\mathrm{R}}^{\text{m}}(C)=\mathcal{C}^{\text{m}}(C):=\sum
\limits_{(\mathbf{x},\mathbf{y})\in\tilde{C}}\mathcal{C}_{\mathbf{xy}},
\end{equation}
where $\mathcal{C}^{\text{m}}(C)$ is the multi-edge capacity of cut $C$. We
may write the following immediate consequence.

\begin{corollary}
Consider a multiple-unicast quantum network $\mathcal{N}$\ with $M$
sender-receiver pairs $(\mathbf{a}_{i},\mathbf{b}_{i})$ communicating via
multipath routing. If the network is distillable, then we may write the
following outer bounds for the capacity region%
\begin{align}
R_{i}^{\text{m}}  &  \leq\min_{C:\mathbf{a}_{i}|\mathbf{b}_{i}}\mathcal{C}%
^{\text{m}}(C)~~\text{for any }i,\\
R_{i}^{\text{m}}+R_{j}^{\text{m}}  &  \leq\min_{C:\mathbf{a}_{i}\mathbf{a}%
_{j}|\mathbf{b}_{i}\mathbf{b}_{j}}\mathcal{C}^{\text{m}}(C)~~\text{for any
}i\neq j\\
&  \vdots\nonumber\\
\sum\limits_{i=1}^{M}R_{i}^{\text{m}}  &  \leq\min_{C:\{\mathbf{a}%
_{i}\}|\{\mathbf{b}_{i}\}}\mathcal{C}^{\text{m}}(C),
\end{align}
where $\mathcal{C}^{\text{m}}(C)$ is the multi-edge capacity of cut $C$.
\end{corollary}

Achievable lower bounds may be determined by combining the point-to-point
composition strategy with classical routing algorithms based on the
maximization of multiple flows. For the specific case $M=1$, the outer bound
is achievable and we retrieve the max-flow min-cut theorem for quantum
communications~\cite{netpaper,longVersion}. For $M>2$, achievable lower bounds
may be found by exploiting classical literature on multicommodity flow
algorithms, e.g., Ref.~\cite{TCHu} which showed a version of the max-flow
min-cut theorem for undirected networks with two commodities, and
Ref.~\cite{Schrijver} which discusses extensions to more than two commodities.

\section{Multicast quantum networks\label{SECmulticastSINGLE}}

Let us now consider a multicast scenario, where Alice $\mathbf{a}$ aims at
simultaneously communicate generally-different messages to a set of
$M$\ receivers, i.e., a set of Bobs $\mathbf{\{b}_{i}\}$. Because of the
implicit parallel nature of this communication process, it is directly
formulated under the assumption of multipath routing. We can easily generalize
the description of the one-sender one-receiver flooding protocol to the
present case of multiple receivers.

In a $1$-to-$M$ multicast network protocol, the quantum network $\mathcal{N}$
is subject to an orientation where Alice is treated as a source, while the
various Bobs are destination points, each one being a receiver but also a
potential relay for another receiver (so that they are not necessarily sinks
in the general case). Each end-to-end simultaneous communication between Alice
and the Bobs consists of a sequence of multicasts from each point of the
network to its out-neighborhood, assisted by network LOCCs. This is done in a
flooding fashion so that each edge of the network is exploited. The
orientation of the network may be updated and optimized at each round of the protocol.

The sequence of orientations and the network LOCCs define the multicast
flooding protocol $\mathcal{P}_{\text{multicast}}^{\text{flood}}$. After $n$
uses of the network, Alice and the $M$ Bobs will share an output state
$\rho_{\mathbf{a\{b}_{i}\}}^{n}$ which is $\varepsilon$-close to a target
state
\begin{equation}
\phi:=%
{\textstyle\bigotimes_{i=1}^{M}}
\phi_{\mathbf{ab}_{i}}^{\otimes nR_{i}^{\varepsilon,n}}.
\end{equation}
where $\phi_{\mathbf{ab}_{i}}$ is a one-bit state (private bit or ebit) for
the pair of points $(\mathbf{a},\mathbf{b}_{i})$ and $nR_{i}^{\varepsilon,n}$
the number of its copies. Note that this is a compact notation which involves
countable sets of systems $\mathbf{a}=(a,a^{\prime},a^{\prime\prime},\ldots)$
and $\mathbf{b}_{i}=(b_{i},b_{i}^{\prime},b_{i}^{\prime\prime},\ldots)$.
Therefore, the tensor product $\phi_{\mathbf{ab}_{1}}^{\otimes nR_{1}%
^{\varepsilon,n}}\otimes\phi_{\mathbf{ab}_{2}}^{\otimes nR_{2}^{\varepsilon
,n}}$ explicitly means $\phi_{ab_{1}}^{\otimes nR_{1}^{\varepsilon,n}}%
\otimes\phi_{a^{\prime}b_{2}^{\prime}}^{\otimes nR_{2}^{\varepsilon,n}}$, so
that there are different systems involved in Alice's side.

By taking the limit of large $n$, small $\varepsilon$, and optimizing over
$\mathcal{P}_{\text{multicast}}^{\text{flood}}$, we define the capacity region
associated with the achievable rates $(R_{1},\ldots,R_{M})$. In particular, we
may define a unique capacity which is associated with the symmetric condition
$R_{1}=\ldots=R_{M}$ (or, more precisely, with a guaranteed common rate $R$
with each Bob). In fact, we may consider a symmetric type of protocol
$\mathcal{\tilde{P}}_{\text{multicast}}^{\text{flood}}$ whose target state
$\phi$ must have $nR_{i}^{\varepsilon,n}\geq nR_{\varepsilon,n}$ bits for any
$i$. Then, by taking the asymptotic limit of large $n$ small $\varepsilon$,
and maximizing over all such protocols, we may define the multicast network
capacity%
\begin{equation}
\mathcal{C}^{M}(\mathcal{N})=\sup_{\mathcal{\tilde{P}}_{\text{multicast}%
}^{\text{flood}}}\lim_{\varepsilon,n}R_{\varepsilon,n}. \label{multiCC}%
\end{equation}
This rate quantifies the maximum number of target bits per network use
(multipath transmission) that Alice may simultaneously share with each Bob in
the destination set $\mathbf{\{b}_{i}\}$. We have the usual hierarchy
$Q_{2}^{M}(\mathcal{N})=D_{2}^{M}(\mathcal{N})\leq K^{M}(\mathcal{N})$ when we
specify the target state. We can now state the following general bound.

\begin{theorem}
[Quantum multicast]\label{TheomultiCAST}Let us consider a multicast quantum
network $\mathcal{N}$ with one sender and $M$ receivers $\mathbf{\{b}_{i}%
\}$.\ Adopt a simulation of the network with a resource representation
$\sigma(\mathcal{N})=\{\sigma_{\mathbf{xy}}\}_{(\mathbf{x},\mathbf{y})\in E}$.
In particular, $\sigma(\mathcal{N})$\ can be a Choi-representation for a
teleportation-covariant $\mathcal{N}$. Then we have the following outer bounds
for the capacity region
\begin{align}
R_{i}  &  \leq E_{\mathrm{R}}^{\text{m}}(i):=\min_{C:\mathbf{a}|\mathbf{b}%
_{i}}E_{\mathrm{R}}^{\text{m}}(C)~~\text{for any }i,\label{cvcc}\\
R_{i}+R_{j}  &  \leq\min_{C:\mathbf{a}|\mathbf{b}_{i}\mathbf{b}_{j}%
}E_{\mathrm{R}}^{\text{m}}(C)~~\text{for any }i\neq j\\
&  \vdots\nonumber\\
\sum\limits_{i=1}^{M}R_{i}  &  \leq\min_{C:\mathbf{a}|\{\mathbf{b}_{i}%
\}}E_{\mathrm{R}}^{\text{m}}(C), \label{cvcc3}%
\end{align}
where $E_{\mathrm{R}}^{\text{m}}(C)$ is the multi-edge flow of REE\ through
cut $C$. In particular, the multicast network capacity satisfies%
\begin{equation}
\mathcal{C}^{M}(\mathcal{N})\leq\min_{i\in\{1,M\}}E_{\mathrm{R}}^{\text{m}%
}(i). \label{CMNmultics}%
\end{equation}
It is understood that formulations may be asymptotic for quantum networks with
bosonic channels.
\end{theorem}

\textbf{Proof.}~~Consider the upper bound given by secret-key generation. With
respect to the bipartition $\mathbf{a}|\{\mathbf{b}_{i}\}$, we may write the
usual steps starting form the distillable key of the target state
\begin{align}
K_{\mathrm{D}}(\mathbf{a}|\{\mathbf{b}_{i}\})_{\phi}  &  =n\sum\limits_{i=1}%
^{M}R_{i}^{\varepsilon,n}\\
&  \leq E_{\mathrm{R}}(\mathbf{a}|\{\mathbf{b}_{i}\})_{\phi}\\
&  \leq E_{\mathrm{R}}(\mathbf{a}|\{\mathbf{b}_{i}\})_{\rho^{n}}%
+\delta(\varepsilon,d),
\end{align}
leading to the asymptotic limit%
\begin{equation}
\lim_{\varepsilon,n}\sum\limits_{i=1}^{M}R_{i}^{\varepsilon,n}\leq
\underset{n\rightarrow+\infty}{\lim}~n^{-1}E_{\mathrm{R}}(\mathbf{a}%
|\{\mathbf{b}_{i}\})_{\rho^{n}}. \label{repkkkj2}%
\end{equation}

For any cut $C:\mathbf{a}|\{\mathbf{b}_{i}\}$ of the (simulated) network, we
may write the decomposition%
\begin{equation}
\rho_{\mathbf{a}\{\mathbf{b}_{i}\}}^{n}(C)=\bar{\Lambda}_{\mathbf{a}%
\{\mathbf{b}_{i}\}}^{C}\left[  \underset{(\mathbf{x},\mathbf{y})\in\tilde{C}}{%
{\textstyle\bigotimes}
}~\sigma_{\mathbf{xy}}^{\otimes n}\right]  ,
\end{equation}
for some trace-preserving LOCC $\bar{\Lambda}_{\mathbf{a}\{\mathbf{b}_{i}%
\}}^{C}$. By replacing $\rho^{n}=\rho_{\mathbf{a}\{\mathbf{b}_{i}\}}^{n}(C)$
in Eq.~(\ref{repkkkj2}), we therefore get%
\begin{equation}
\lim_{\varepsilon,n}\sum\limits_{i=1}^{M}R_{i}^{\varepsilon,n}\leq
\sum\limits_{(\mathbf{x},\mathbf{y})\in\tilde{C}}E_{\mathrm{R}}(\sigma
_{\mathbf{xy}}):=E_{\mathrm{R}}^{\text{m}}(C).
\end{equation}
By minimizing over the cuts and maximizing over the protocols, we may write%
\begin{equation}
\sum\limits_{i=1}^{M}R_{i}\leq\min_{C:\mathbf{a}|\{\mathbf{b}_{i}%
\}}E_{\mathrm{R}}^{\text{m}}(C).
\end{equation}

The other conditions in Eqs.~(\ref{cvcc})-(\ref{cvcc3}) are obtained by
setting part of the rates $R_{i}^{\varepsilon,n}$ to zero in the target state
(as in the previous proofs). In particular, set $R_{i}^{\varepsilon,n}\neq0$
for some $i$, while $R_{j}^{\varepsilon,n}=0$ for any $j\neq i$. The target
state becomes $\phi:=\phi_{\mathbf{ab}_{i}}^{\otimes nR_{i}^{\varepsilon,n}%
}\otimes\sigma_{\text{sep}}$ and we repeat the derivation with respect to the
bipartition $\mathbf{a|b}_{i}$. This leads to
\begin{equation}
\lim_{\varepsilon,n}R_{i}^{\varepsilon,n}\leq\underset{n\rightarrow+\infty
}{\lim}~n^{-1}E_{\mathrm{R}}(\mathbf{a}|\mathbf{b}_{i})_{\rho^{n}},
\label{multi11}%
\end{equation}
where we may directly consider the reduced state
\begin{equation}
\rho^{n}=\rho_{\mathbf{ab}_{i}}^{n}=\mathrm{Tr}_{\{\mathbf{b}_{j\neq i}%
\}}\left[  \rho_{\mathbf{a}\{\mathbf{b}_{1},\ldots,\mathbf{b}_{M}\}}%
^{n}\right]  .
\end{equation}

For any cut $C:\mathbf{a}|\mathbf{b}_{i}$, we therefore have%
\begin{equation}
\rho_{\mathbf{ab}_{i}}^{n}(C)=\bar{\Lambda}_{\mathbf{ab}_{i}}^{C}\left[
\underset{(\mathbf{x},\mathbf{y})\in\tilde{C}}{%
{\textstyle\bigotimes}
}~\sigma_{\mathbf{xy}}^{\otimes n}\right]  , \label{multi22}%
\end{equation}
which leads to $\lim_{\varepsilon,n}R_{i}^{\varepsilon,n}\leq E_{\mathrm{R}%
}^{\text{m}}(C)$. By minimizing over the cuts, one gets
\begin{equation}
\lim_{\varepsilon,n}R_{i}^{\varepsilon,n}\leq E_{\mathrm{R}}^{\text{m}%
}(i):=\min_{C:\mathbf{a}|\mathbf{b}_{i}}E_{\mathrm{R}}^{\text{m}}(C).
\label{bbbmmm}%
\end{equation}
Since this is true for any protocol $\mathcal{P}^{M}$, it can be extended to
the achievable rates, i.e., we get Eq.~(\ref{cvcc}).

For the multicast network capacity, just note that
\begin{equation}
\lim_{\varepsilon,n}R^{\varepsilon,n}\leq\min_{i}\{\lim_{\varepsilon,n}%
R_{i}^{\varepsilon,n}\}.
\end{equation}
Therefore, from Eq.~(\ref{bbbmmm}), we may write%
\begin{equation}
\lim_{\varepsilon,n}R^{\varepsilon,n}\leq\min_{i}E_{\mathrm{R}}^{\text{m}}(i).
\end{equation}
This is true for any symmetric protocol $\mathcal{P}_{\text{sym}}^{M}$ which
leads to the result of Eq.~(\ref{CMNmultics}). Results are extended to
asymptotic simulations of bosonic channels in the usual way.~$\blacksquare$

As usual, in the case of distillable networks, we may prove stronger results.
As a direct consequence of Theorem~\ref{TheomultiCAST}, we may write the
following cutset bound.

\begin{corollary}
Consider a multicast quantum network $\mathcal{N}$ with one sender and $M$
receivers $\mathbf{\{b}_{i}\}$.\ If the network is distillable, then we have
the following outer bounds for the capacity region
\begin{align}
R_{i}  &  \leq\mathcal{C}^{\text{m}}(i)=\min_{C:\mathbf{a}|\mathbf{b}_{i}%
}\mathcal{C}^{\text{m}}(C)~~\text{for any }i,\\
R_{i}+R_{j}  &  \leq\min_{C:\mathbf{a}|\mathbf{b}_{i}\mathbf{b}_{j}%
}\mathcal{C}^{\text{m}}(C)~~\text{for any }i\neq j\\
&  \vdots\nonumber\\
\sum\limits_{i=1}^{M}R_{i}  &  \leq\min_{C:\mathbf{a}|\{\mathbf{b}_{i}%
\}}\mathcal{C}^{\text{m}}(C),
\end{align}
where $\mathcal{C}^{\text{m}}(C)$ is the multi-edge capacity of cut $C$ and
$\mathcal{C}^{\text{m}}(i)$ is the multipath capacity between the sender and
the $i$th receiver (in a unicast setting). In particular, the multicast
network capacity must satisfy the bound%
\begin{equation}
\mathcal{C}^{M}(\mathcal{N})\leq\min_{i\in\{1,M\}}\mathcal{C}^{\text{m}}(i)~.
\label{cutsetDIST}%
\end{equation}

\end{corollary}

Our previous results refer to the general case of multiple independent
messages. In a multicast quantum network, this means that Alice distributes
$M$ different sequences of\ target bits to the $M$ Bobs $\{\mathbf{b}_{i}\}$.
For instance, these may represent $M$ different secret keys, one for each Bob
in the destination set. For this specific task (key distribution), the
multicast capacity of the network $\mathcal{C}^{M}(\mathcal{N})$ becomes a
multicast secret-key capacity $\mathcal{K}^{M}(\mathcal{N})$.

In QKD, it is interesting to consider the variant scenario where
Alice distributes exactly the same secret key to all Bobs
$\{\mathbf{b}_{i}\}$, e.g., to enable a quantum-secured conference
among these parties. For this particular task, we may define a
single-key version of the multicast secret-key capacity, that we
denote as $\mathcal{K}_{\text{1-key}}^{M}(\mathcal{N})$. This
represents the maximum rate at which Alice may distribute the same
secret key to all Bobs in each parallel use of the network. It is
clear that we have
$\mathcal{K}_{\text{1-key}}^{M}(\mathcal{N})\geq\mathcal{K}^{M}(\mathcal{N})$,
just because Alice may use $M-1$ distributed keys to encrypt and
share the shortest key with all Bobs.

\section{Multiple-multicast quantum networks\label{SECmulticastMANY}}

In the multiple-multicast quantum network, we have $M_{A}$ Alices
$\{\mathbf{a}_{1},\ldots,\mathbf{a}_{i},\ldots,\mathbf{a}_{M_{A}}\}$, each of
them communicating generally-different messages with a destination set of
$M_{B}$ Bobs $\{\mathbf{b}_{1},\ldots,\mathbf{b}_{j},\ldots,\mathbf{b}_{M_{B}%
}\}$ by means of multipath routing. Each end-to-multiend multicast
$\mathbf{a}_{i}\rightarrow\{\mathbf{b}_{j}\}$ is associated with the
distribution of $M_{B}$ independent sequences of target bits (e.g., secret
keys) between the $i$th Alice $\mathbf{a}_{i}$ and each Bob $\mathbf{b}_{j}$
in the destination set. The description of a multiple-multicast protocol for a
quantum network follows the same main features discussed for the case of a
single-multicast network ($M_{A}=1$). Because we have multiple senders and
receivers, here we need to consider all possible orientations of the network.
Each use of the quantum network is performed under some orientation which is
adopted by the points for their out-neighborhood multicasts, suitably assisted
by network LOCCs. Use after use, these steps define a multiple-multicast
flooding protocol $\mathcal{P}_{\text{M-multicast}}^{\text{flood}}$.

After $n$ uses, the ensembles of Alices and Bobs share an output state
$\rho_{\{\mathbf{a}_{i}\}\mathbf{\{b}_{j}\}}^{n}$ which is $\varepsilon$-close
to a target state
\begin{equation}
\phi:=%
{\textstyle\bigotimes_{i=1}^{M_{A}}}
{\textstyle\bigotimes_{j=1}^{M_{B}}}
\phi_{\mathbf{a}_{i}\mathbf{b}_{j}}^{\otimes nR_{ij}^{\varepsilon,n}}.
\end{equation}
where $\phi_{\mathbf{a}_{i}\mathbf{b}_{j}}$ is a one-bit state (private bit or
ebit) for the pair $(\mathbf{a}_{i},\mathbf{b}_{j})$ and $nR_{ij}%
^{\varepsilon,n}$ the number of its copies. By taking the limit of large $n$,
small $\varepsilon$, and optimizing over $\mathcal{P}_{\text{M-multicast}%
}^{\text{flood}}$, we define the capacity region for the achievable rates
$\{R_{ij}\}$. Assume the symmetric case where the $i$th Alice $\mathbf{a}_{i}$
achieves the same rate $R_{i1}=\ldots=R_{iM_{B}}$ with all Bobs $\{\mathbf{b}%
_{j}\}$ (or, more precisely, a guaranteed common rate $R_{i}$). This means to
consider symmetric protocols whose target state $\phi$ must have $\min
_{j}R_{ij}^{\varepsilon,n}\geq R_{i}^{\varepsilon,n}$ bits for any $i$. By
taking the asymptotic limit of $R_{i}^{\varepsilon,n}$ for large $n$, small
$\varepsilon$, and maximizing over all these symmetric protocols, we may
define the capacity region for the achievable multicast rates $(R_{1}%
,\ldots,R_{M_{A}})$. In the latter set, rate $R_{i}$ provides the minimum
number of target bits per use that the $i$th Alice may share with each Bob in
the destination set $\mathbf{\{b}_{j}\}$ (in the multi-message setting, i.e.,
assuming independent sequences shared with the various Bobs). We have the
following outer bounds to the capacity region.

\begin{theorem}
[Quantum multiple-multicast]\label{Theomultimulti}Let us consider a
multiple-multicast quantum network $\mathcal{N}=(P,E)$ where each of the
$M_{A}$ senders $\mathbf{\{a}_{i}\}$ communicates with $M_{B}$ receivers
$\mathbf{\{b}_{j}\}$ at the multicast rate $R_{i}$. Adopt a simulation of
$\mathcal{N}$ with some resource representation $\sigma(\mathcal{N}%
)=\{\sigma_{\mathbf{xy}}\}_{(\mathbf{x},\mathbf{y})\in E}$, which may be a
Choi-representation for a teleportation-covariant $\mathcal{N}$. Then, we have
the following outer bounds for the capacity region%
\begin{align}
R_{i}  &  \leq\min_{\substack{C:\mathbf{a}_{i}\in\mathbf{A}\\\{\mathbf{b}%
_{j}\}\cap\mathbf{B}\neq\emptyset}}E_{\mathrm{R}}^{\text{m}}%
(C),\label{setCVB0}\\
R_{i}+R_{j}  &  \leq\min_{\substack{C:\mathbf{a}_{i},\mathbf{a}_{j}%
\in\mathbf{A}\\\{\mathbf{b}_{j}\}\cap\mathbf{B}\neq\emptyset}}E_{\mathrm{R}%
}^{\text{m}}(C),\\
&  \vdots\nonumber\\
\sum\limits_{i=1}^{M_{A}}R_{i}  &  \leq\min_{\substack{C:\{\mathbf{a}%
_{i}\}\subseteq\mathbf{A}\\\{\mathbf{b}_{j}\}\cap\mathbf{B}\neq\emptyset
}}E_{\mathrm{R}}^{\text{m}}(C), \label{setCVB}%
\end{align}
where $E_{\mathrm{R}}^{\text{m}}(C)$ is the multi-edge flow of REE\ through
cut $C$. For a distillable network, we may write the bounds in
Eqs.~(\ref{setCVB0})-(\ref{setCVB}) with $E_{\mathrm{R}}^{\text{m}%
}(C)=\mathcal{C}^{\text{m}}(C)$, i.e., in terms of the multi-edge capacity of
the cuts.
\end{theorem}

\textbf{Proof.}~~Consider the upper bound given by secret-key generation. With
respect to the bipartition $\{\mathbf{a}_{i}\}|\{\mathbf{b}_{j}\}$, we can
manipulate the distillable key $K_{\mathrm{D}}$ of the target state $\phi$ as
follows
\begin{align}
K_{\mathrm{D}}(\{\mathbf{a}_{i}\}|\{\mathbf{b}_{j}\})_{\phi}  &
=n\sum\limits_{i=1}^{M_{A}}\sum\limits_{j=1}^{M_{B}}R_{ij}^{\varepsilon,n}\\
&  \leq E_{\mathrm{R}}(\{\mathbf{a}_{i}\}|\{\mathbf{b}_{j}\})_{\phi}\\
&  \leq E_{\mathrm{R}}(\{\mathbf{a}_{i}\}|\{\mathbf{b}_{j}\})_{\rho^{n}%
}+\delta(\varepsilon,d),
\end{align}
leading to the asymptotic limit%
\begin{equation}
\lim_{\varepsilon,n}\sum\limits_{i=1}^{M_{A}}\sum\limits_{j=1}^{M_{B}}%
R_{ij}^{\varepsilon,n}\leq\underset{n\rightarrow+\infty}{\lim}~n^{-1}%
E_{\mathrm{R}}(\{\mathbf{a}_{i}\}|\{\mathbf{b}_{j}\})_{\rho^{n}}.
\label{repkkkj3}%
\end{equation}

For any cut $C:\{\mathbf{a}_{i}\}|\{\mathbf{b}_{j}\}$ of the (simulated)
network, we may write the decomposition%
\begin{equation}
\rho_{\{\mathbf{a}_{i}\}\{\mathbf{b}_{j}\}}^{n}(C)=\bar{\Lambda}%
_{\{\mathbf{a}_{i}\}\{\mathbf{b}_{j}\}}^{C}\left[  \underset{(\mathbf{x}%
,\mathbf{y})\in\tilde{C}}{%
{\textstyle\bigotimes}
}\sigma_{\mathbf{xy}}^{\otimes n}\right]  ,
\end{equation}
and manipulate Eq.~(\ref{repkkkj3}) into the following%
\begin{equation}
\lim_{\varepsilon,n}\sum\limits_{i=1}^{M_{A}}\sum\limits_{j=1}^{M_{B}}%
R_{ij}^{\varepsilon,n}\leq\sum\limits_{(\mathbf{x},\mathbf{y})\in\tilde{C}%
}E_{\mathrm{R}}(\sigma_{\mathbf{xy}}):=E_{\mathrm{R}}^{\text{m}}(C).
\end{equation}
By minimizing over the cuts and maximizing over the protocols, we may write%
\begin{equation}
\sum\limits_{i=1}^{M_{A}}\sum\limits_{j=1}^{M_{B}}R_{ij}\leq\min
_{C:\{\mathbf{a}_{i}\}|\{\mathbf{b}_{j}\}}E_{\mathrm{R}}^{\text{m}}(C)~.
\end{equation}

By setting part of the rates $R_{ij}^{\varepsilon,n}$ to zero in the target
state, we derive the full set of conditions
\begin{align}
\sum\limits_{i=1}^{M_{A}}\sum\limits_{j=1}^{M_{B}}R_{ij}  &  \leq
\min_{C:\{\mathbf{a}_{i}\}|\{\mathbf{b}_{j}\}}E_{\mathrm{R}}^{\text{m}%
}(C),\label{xzx1}\\
&  \vdots\nonumber\\
R_{ij}+R_{kl}  &  \leq\min_{C:\mathbf{a}_{i}\mathbf{a}_{k}|\mathbf{b}%
_{j}\mathbf{b}_{l}}E_{\mathrm{R}}^{\text{m}}(C),\label{xzx2}\\
R_{ij}  &  \leq\min_{C:\mathbf{a}_{i}|\mathbf{b}_{j}}E_{\mathrm{R}}^{\text{m}%
}(C). \label{xzx3}%
\end{align}

The latter conditions are valid for the end-to-end rates $R_{ij}$ achievable
between each pair $(\mathbf{a}_{i},\mathbf{b}_{j})$. We are interested in the
achievable multicast rates $\{R_{i}\}$ between each sender $\mathbf{a}_{i}$
and all receivers $\{\mathbf{b}_{j}\}$. Corresponding conditions can be
derived by considering a subset of protocols with target state of the type
\begin{equation}
\phi_{k}:=%
{\textstyle\bigotimes_{i=1}^{M_{A}}}
\phi_{\mathbf{a}_{i}\mathbf{b}_{k}}^{\otimes nR_{ik}^{\varepsilon,n}}%
\otimes\sigma_{\text{sep}},
\end{equation}
for some $k$, where all Alices $\{\mathbf{a}_{i}\}$\ aim to optimize their
rates $\{R_{ik}^{\varepsilon,n}\}$ with some fixed Bob $\mathbf{b}_{k}$, so
that $R_{ij}^{\varepsilon,n}=0$ for any $j\neq k$. By repeating the previous
steps with respect to the bipartition $\{\mathbf{a}_{i}\}|\mathbf{b}_{k}$, we
obtain%
\begin{equation}
\lim_{\varepsilon,n}\sum\limits_{i=1}^{M_{A}}R_{ik}^{\varepsilon,n}\leq
\min_{C:\{\mathbf{a}_{i}\}|\mathbf{b}_{k}}E_{\mathrm{R}}^{\text{m}}(C).
\end{equation}
Since we have $R_{i}^{\varepsilon,n}\leq\min_{j}R_{ij}^{\varepsilon,n}\leq
R_{ik}^{\varepsilon,n}$ for any $i$, we can then write the same inequality for
$\lim_{\varepsilon,n}\sum\nolimits_{i=1}^{M_{A}}R_{i}^{\varepsilon,n}$. Then,
by optimizing over the protocols, we get%
\begin{equation}
\sum\limits_{i=1}^{M_{A}}R_{i}\leq\min_{C:\{\mathbf{a}_{i}\}|\mathbf{b}_{k}%
}E_{\mathrm{R}}^{\text{m}}(C).
\end{equation}
Because we may repeat the previous reasoning for any $k$, we may write%
\begin{equation}
\sum\limits_{i=1}^{M_{A}}R_{i}\leq\min_{C}E_{\mathrm{R}}^{\text{m}}(C),
\label{yyu1}%
\end{equation}
with $C=(\mathbf{A},\mathbf{B})$ such that $\{\mathbf{a}_{i}\}\subseteq
\mathbf{A}$ and $\{\mathbf{b}_{j}\}\cap\mathbf{B}\neq\emptyset$.

Now, for any fixed $k$, impose that the rates $\{R_{ik}^{\varepsilon,n}\}$ are
zero for some of the Alices $\{\mathbf{a}_{i}\}$. If we set $R_{ik}%
^{\varepsilon,n}\neq0$ for a pair $(\mathbf{a}_{i},\mathbf{b}_{k})$, then the
condition $R_{i}^{\varepsilon,n}\leq R_{ik}^{\varepsilon,n}$ leads to%
\begin{equation}
R_{i}\leq\min_{C:\mathbf{a}_{i}|\mathbf{b}_{k}}E_{\mathrm{R}}^{\text{m}}(C).
\end{equation}
Because the reasoning can be repeated for any $k$, we may then write%
\begin{equation}
R_{i}\leq\min_{C}E_{\mathrm{R}}^{\text{m}}(C), \label{yyu2}%
\end{equation}
with $C=(\mathbf{A},\mathbf{B})$ such that $\mathbf{a}_{i}\in\mathbf{A}$ and
$\{\mathbf{b}_{j}\}\cap\mathbf{B}\neq\emptyset$. Extending the previous
reasoning to two non-zero rates $R_{ik}^{\varepsilon,n}\neq0$ and
$R_{jk}^{\varepsilon,n}\neq0$ leads to%
\begin{equation}
R_{i}+R_{j}\leq\min_{C}E_{\mathrm{R}}^{\text{m}}(C), \label{yyu3}%
\end{equation}
with $C=(\mathbf{A},\mathbf{B})$ such that $\mathbf{a}_{i},\mathbf{a}_{j}%
\in\mathbf{A}$ and $\{\mathbf{b}_{j}\}\cap\mathbf{B}\neq\emptyset$. Other
similar conditions can be derived for the multicast rates, so that we get the
result of Eqs.~(\ref{setCVB0})-(\ref{setCVB}). Finally, for a distillable
network we have $E_{\mathrm{R}}^{\text{m}}(C)=\mathcal{C}^{\text{m}}(C)$ and,
therefore, it is immediate to express these results in terms of the multi-edge
capacities of the cuts.~$\blacksquare$

\section{Conclusions\label{SECconclusions}}

In this information-theoretic work, we have investigated the
ultimate rates for transmitting quantum information, distributing
entanglement and generating secret keys between multiple senders
and receivers in an arbitrary quantum network, assuming single- or
multi-path routing strategies. We have established general
single-letter REE upper bounds for the various multiend capacities
associated to the various configurations of multiple-unicast,
multicast, and multiple-multicast quantum networks. The bounds
apply to networks connected by arbitrary quantum channels (at any
dimension) with more specific formulations in the case of
teleportation-covariant and distillable channels. In particular,
the case of quantum networks connected by bosonic channels is
implicitly treated by using asymptotic LOCC simulations, so that
the results are automatically proven for fundamental noise models
at the optical and telecom regimes, such as the pure-loss
channels.

The present paper provides a multiend generalization of the results of
Ref.~\cite{netpaper} (first appeared in Ref.~\cite{longVersion}) for the basic
end-to-end (unicast) scenario. It also extends the results presented in
Ref.~\cite{Ric1}\ from single-hop to multi-hop quantum networks. Due to the
much more complex scenarios associated with the simultaneous multi-hop quantum
communication among multiple senders and receivers, we could only bound the
capacity regions in the various configurations analyzed, so that their full
characterization remains an open question for further investigations.

\bigskip

\textbf{Acknowledgments}.~This work has been supported by the EPSRC\ via the
`UK Quantum Communications HUB' (EP/M013472/1) and by the European Union via
the project `Continuous Variable Quantum Communications' (CiViQ, no 820466).

\end{document}